# Automated Classification of Hand-grip action on Objects using Machine Learning

ANJU MISHRA, Amity University Uttar Pradesh
SHANU SHARMA, Amity University Uttar Pradesh
SANJAY KUMAR, Oxford Brooks University
PRIYA RANJAN, Amity University Uttar Pradesh
AMIT UJLAYAN, Gautam Buddha University

Brain computer interface is the current area of research to provide assistance to disabled persons. To cope up with the growing needs of BCI applications, this paper presents an automated classification scheme for handgrip actions on objects by using Electroencephalography (EEG) data. The presented approach focuses on investigation of classifying correct and incorrect handgrip responses for objects by using EEG recorded patterns. The method starts with preprocessing of data, followed by extraction of relevant features from the epoch data in the form of discrete wavelet transform (DWT), and entropy measures. After computing feature vectors, artificial neural network classifiers used to classify the patterns into correct and incorrect handgrips on different objects. The proposed method was tested on real dataset, which contains EEG recordings from 14 persons. The results showed that the proposed approach is effective and may be useful to develop a variety of BCI based devices to control hand movements.

## 1 INTRODUCTION

Brain Computer Interface (BCI) is an emerging area of study, which makes it possible to act as a communication medium between a human and the computer. There are numerous possibilities and application areas for BCIs as they provide an easy interface to understand the functioning of human brain. BCI has applications in medical, industrial, experimental psychology and neurorehabilitation to name a few. The most effective way to understand the functioning of brain is by analyzing electroencephalogram recordings, commonly known as EEG data. The EEG recordings contain cortical potentials, which occur during various mental processes. These signals comprises of different frequency sub-bands: delta (4 Hz), theta (4-7 Hz), alpha or mu (8-12 Hz), beta (12-30), and gamma (30-100 Hz) bands, to provide for ease of analysis. Studies show that mu rhythm is more sensitive to correct and incorrect hand grip and responds strongly especially over motor and pre-motor cortex area of brain. Particularly the event related desynchronization of mu rhythm found to be more profound for congruent grip on object rather than for incongruent grip response. Here, we present a fully automated system capable of sensing the correctness of grip response over familiar objects by analyzing EEG data. The system starts by taking EEG data and then applying a pre-processing step over recorded data. This pre-processing step consists of re-referencing of EEG data over extra electrodes using EEG Lab in MATLAB environment. The electrodes chosen for this task was electrode numbers 129 and 130. After this, epoch extraction was performed on the re-referenced data and resulted in a total of 89 epochs. The epoched EEG signals were filtered using a band pass filter to isolate mu rhythm (8-12 Hz). Finally, features were extracted form mu rhythm data and passed on to a neural network classifier for training. The system can be used for training purposes in neurorehabilitation of disabled persons and can be used to develop a variety of BCI based devices to control limb movements in robotic and prosthetic settings as well.

A neural-network-based automated epileptic detection system from EEG data was proposed in [1]. The system used an approximate entropy (ApEn) as the input feature. ApEn is a statistical measures that calculates the predictability of the current amplitude values of a physiological signal based on its previous amplitude values. The authors found that the value of the ApEn drops sharply during an epileptic seizure. Two different types of neural networks, Elman and probabilistic neural networks, were considered for implementing the system. In another work a multilayer perceptron neural network (MLPNN) based classification model was proposed as a decision support mechanism in the epilepsy treatment. [2] EEG signals were decomposed into frequency sub-bands using discrete wavelet transform (DWT) and then wavelet coefficients were clustered using the K-means clustering algorithm. The probability distributions were computed for different clusters, and then used as inputs to the MLPNN model. In [3] a method was proposed for automatic detection of normal, pre-ictal, and ictal conditions from recorded EEG signals. Four entropy features: Approximate Entropy (ApEn), Sample Entropy (SampEn), Phase Entropy 1 (S1), and Phase Entropy 2 (S2) were used as features and seven different classifiers: Fuzzy Sugeno Classifier (FSC), Support Vector Machine (SVM), K-Nearest Neighbour (KNN), Probabilistic Neural Network (PNN), Decision Tree (DT), Gaussian Mixture Model (GMM), and Naive Bayes Classifier (NBC) were used to assess the performance of classifier for EEG data. In [4] an EEG based classification system was proposed for classifying driver fatigue versus alert state. The system used independent component by entropy rate bound minimization analysis (ERBM-ICA) for the source separation, autoregressive (AR) modeling for the features extraction, and a Bayesian neural network as classification algorithm. In [5] a BCI system was presented based EEG signals obtained from five mental tasks (baseline, math, mental letter composing, geometric figure rotation and visual counting). For feature extraction Wavelet Transform (WT), Fast Fourier Transform (FFT) and Principal Component Analysis (PCA) were used. Artificial Neural Network (ANN) and Support Vector Machines (SVMs) classifiers were used for classifying different combinations of mental tasks. The authors of [11] proposed a method for EEG-based automated diagnosis of epilepsy. The method involved detection of key points at multiple scales in EEG signals using a pyramid of difference of Gaussian filtered signals then Local binary patterns (LBPs) were calculated at these key points and the histogram of these patterns are considered as the feature and finally fed to a SVM classifier for the classification. In [12] online cognitive failures in driving was assessed EEG signals. Visual alertness of the driver is detected by classifying EEG signals into alert and non-alert states. A type-2 fuzzy set induced neural classifier was used to eliminate the uncertainty in classification of motor planning. In [13] a new computer aided diagnosis (CAD) of autism based on electroencephalography (EEG) signal analysis was investigated. The authors have used discrete wavelet transform (DWT), entropy (En), and artificial neural network (ANN) to classify person as autistic or healthy. [14,15] have shown the effect of hand grip on object recognition by studying the modulation of the mu rhythm when participants made object decisions to objects and non-objects shown with congruent or incongruent hand-grip actions. ERD is used as an index of neural excitation It has been observed that there is an increased ERD activity in the mu frequency band over scalp motor regions when participants made object decisions to congruently and incongruently gripped objects.

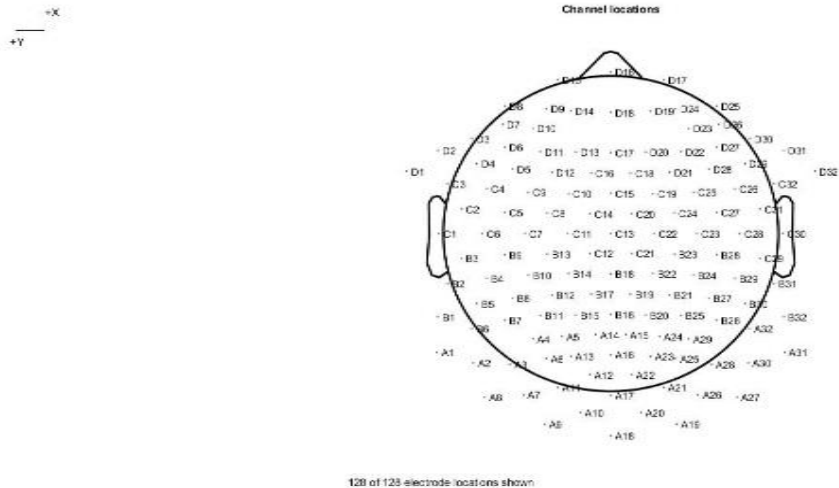

Data Set

The proposed method is tested on real dataset[14,15] which contains EEG recordings from a group of 14 people, out of which 3 were male and 11 were female. The participants were shown images of object with congruent grip, objects with incongruent grip and no grip. Similarly, they were shown images for non-objects with congruent, incongruent and no grip. The task was to decide quickly whether the object was a real one or a non-object. The participants received a total of 180 stimuli of which 30 stimuli pertaining to objects and 30 were for non-object categories. Further, for each object and non-object category, three conditions were chosen: congruent grip, incongruent grip, and no handgrip. Before each task, the participants received twelve practice trials each of which began with the presentation of a fixation point for 1000 ms followed by a target stimulus for 1000 ms. Participants were asked to make a response within 4000 ms after stimulus onset. Here, for the sake of simplicity, we have chosen the data associated with object category only and out of three criteria only the congruent and incongruent cases have been considered. The EEG data was recorded continuously with Ag/AgCl electrodes placed on 128 scalp locations. The electrodes were placed according to 10-5 electrode system. Extra electrodes were used as references and ground. The signals were amplified and sampled at a rate of 1024 Hz using BioSemi Active-Two amplifiers.

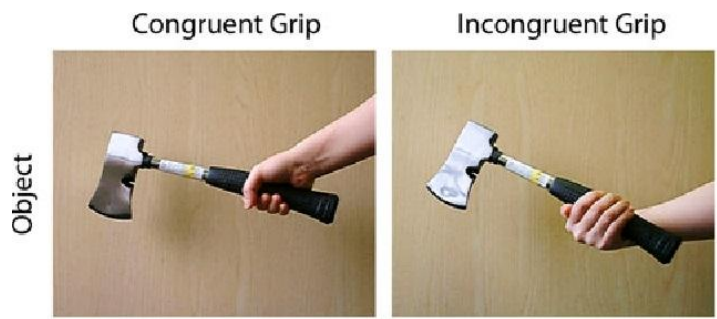

## 2 EXPERIMENTAL AND COMPUTATIONAL DETAILS

The EEG recordings contain cortical potentials which occur during various mental processes. The EEG signals are generally divided into different frequency bands: delta (4 Hz), theta (4-7 Hz), alpha or mu (8-12 Hz), beta (12-30), and gamma (30-100 Hz) bands, to facilitate ease of analysis. The present work features an automated model for analyzing event-related de-synchronization/synchronization (ERD/ERS) of the EEG responses to objects shown with congruent or incongruent grips.

### 2.1 Data Pre-Processing

The EEG continuous data was preprocessed using EEGLAB. This encompasses setting the reference channel indices to 129 and 130, which represents left and right mastoids respectively. Next step is to select first 128 channels from the EEG data as 129 to 134 channels are for referencing and other purpose. After this step, the data are re-referenced by compute average reference function available in EEGLAB. This re-referencing helps in reducing the noise content present in EEG signals. This noise reduction step is viable for correct identification of signal emitting area i.e., the source of signal which in our case is the brain region. The presence of noise is due to many factors such as heartbeats, eye flickers etc occurring while recording the data with EEG electrodes.

The data contains 18 triggers associated with different events. First three triggers correspond to events associated with congruent, incongruent and no grip responses over object category. While next three triggers are associated with congruent, incongruent and no grip responses over non-object category. The rest of the triggers are for reaction time responses. For this study, we have considered the first two triggers associated with object category only as our goal is to classify congruent and incongruent hand grip responses over objects for assessing object affordance form EEG signals. After event selection for epoch extraction, baseline removal was done in EEGLAB. The EEG data so obtained was saved for further processing.

## 2.2 Channel Selection and Rhythm Isolation

Not all channels contribute to deciding that the grip response is correct or incorrect for a gripped object. Studies have shown that different brain areas respond to different types of activities. Therefore, while analyzing the EEG data we only considered EEG data recorded from occipital, motor and parietal areas of brain as most of the motor related activities are controlled from these areas of brain. For each frequency sub-band there is a set of electrodes that responds well to achieve better classification results and these electrodes are associated with specific brain regions. It is also found from preliminary analysis that only certain frequency sub-bands of an EEG signal are important for analyzing certain task. For handgrip responses Sanjay et. al.[14,15] have shown that, out of the five sub-bands only the alpha band responds well for recognizing handgrip actions on objects. This initial investigation suggests several hints to improve the classification task. For our study, we have taken 8 electrodes each from the three areas of interest of brain i.e., occipital, parietal and motor areas, a total of 24 electrodes. These 8 electrodes from each region are further divided into two sets of four electrodes. Four electrodes are selected from left motor area viz., C1, C3, FC1, FC3 and four electrodes are selected from right motor region viz., C2, C4, FC2, FC4 from right brain hemisphere to assess modulation in motor cortex activity. Similarly, two sets of four electrodes viz., PO7, PO5h, POO9h, O1 and PO8, PO6h, POO10h, O2 are taken from left and right occipital areas of brain respectively and P1, P3, PPO3h, PPO5h and P2, P4, PPO4h, PPO6h are selected from left and right parietal brain regions respectively.

Continuous EEG was segmented in epochs from 1000 ms before target-onset to 1000 ms after target-onset. Activity for 1000 ms pre-stimulus was taken as the reference interval. Epochs were discarded if the voltage exceeded ±100μ volt. The remaining epochs were band pass filtered in frequency band of 8–10 Hz and 10–12 Hz for further analysis. For this study alpha band with 8-12 Hz frequency was selected which may reflect focused specific movement activities. Epochs were extracted for each congruent grip and incongruent grip events. Bandpass filtered epoch amplitudes were squared and averaged across all trials for each condition separately. Followed by a Smoothing step on signal traces obtained from previous step. This smoothing was performed by using a moving average window with 100 steps.

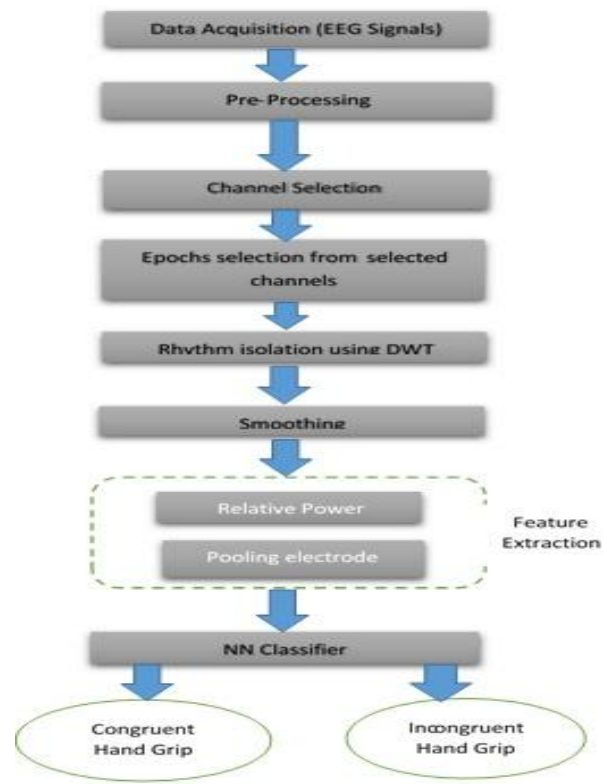

Fig. 1 Basic Flow Diagram of proposed system

## 2.3 Wavelet Decomposition

Wavelet analysis was done to decompose the EEG signal into different frequency bands. Wavelet transform is widely used for time frequency decomposition of acquired signals. Wavelet analysis falls into two categories:

- Continuous Wavelet Transform (CWT)
- Discrete Wavelet Transform (DWT)

CWT is useful for extracting event related potential (ERP) time-frequency features on nonstationary EEG signals and are suitable for effective feature selection which in turn results in significant classification accuracy. However, the drawback of using CWT is it involves an excessive amount of calculations [13]. Therefore, from computational point of view DWT provides for faster calculation of constituting frequency bands. This work uses DWT to decompose the EEG signal into different frequency bands. The two most important factors while performing DWT analysis are selection of decomposition levels and type of mother wavelet used. At each level i the DWT output two types of coefficients: ith level approximation coefficients $A_i$ and detailed coefficients $D_i$ at this level. The input EEG signal is decomposed into a first level approximation A1 and corresponding detailed coefficients D1. Then, the approximation coefficients are decomposed further into second level of approximation and detailed coefficients and so on. Fig 2. The proposed work uses 8-level DWT decomposition with Daubechies-eight (db8) mother wavelet and five EEG sub bands (rhythms) were extracted for analysis purpose. The motive behind extracting sub bands from an EEG band is because of the fact that specific rhythms show strong responses for specific type of events. It has been found that alpha rhythm responds well for motor cortex activations [14, 15]. Therefore, alpha band was the focus of our study for hand grip actions.

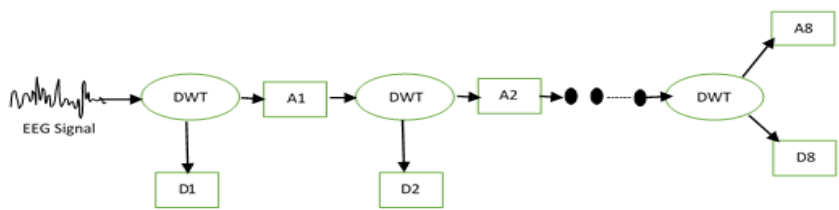

Fig. 2 EEG signal decomposition using 8-level DWT

The frequency bands of EEG signal for 8-level DWT are shown in Table 1. Each sub band, delta (4 Hz), theta (4-7 Hz), alpha or mu (8-12 Hz), beta (12-30), and gamma (30-100 Hz) contributes to some specific characteristic of time series EEG signal.

Table 1. Frequency Sub-bands for Wavelet Coefficients

| Frequency (Hz) | EEG sub-band | Wavelet Coefficient |
| --- | --- | --- |
| 0-4 | Delta | A8 |
| 4-7 | Theta | D8 |
| 8-12 | Alpha | D7 |
| 12-30 | Beta | D6 |
| 30-100 | Gamma | D5 |
| Above 100 | --- | D1 to D4 |

## 2.4 Feature Extraction

Smoothing of sub band (in our case alpha band) was done prior to feature extraction to normalize the coefficients. Next, features were extracted from detailed and approximation coefficients at different levels of DWT transform. Previous studies suggest that a variety of features can be extracted from time series EEG signals viz. power, entropy and statistical features like mean, standard deviation, kurtosis etc. Majority of previous studies have suggested the effectiveness of using entropy and power as major features for analyzing EEG signal data.[16,17,18] The fact that entropy as a feature is strong enough for measuring complexity and regularity of time series data, makes it attractive to use for EEG data analysis. This motivated us to further investigate entropy as a feature modality to analyze hand grip responses for object affordance. Different types of entropy functions exist in literature e.g., log energy entropy function, Shannon entropy, Renyi entropy and threshold entropy function. For our study, we have accounted for Shannon entropy function over EEG sub band.

Shannon entropy developed by Shannon[19] is used to expect the average information value contained in signal and to measure the uncertainty of discrete signal. For time series data X = [x1,x2,….,xn] , Shannon entropy can be calculated as:

$$H = -\sum_{i=1}^{n}(p_i)log_2(p_i) \tag{1}$$

Where n is the number of data points in X and $p_i$ is the normalized probability for each xi. The next step after feature extraction was to classify EEG signal for congruent and incongruent grip on objects. For this, the feature vectors used to train an artificial neural network for classification purpose.

## 2.5 Artificial Neural Network Classifier

Artificial neural network (ANN) is a widely adopted approach in the biomedical domain for classifying data. ANN is an information processing system which imitates human cognitive processing. ANN is a connected network of several neural computational units called artificial neurons. Each artificial neuron is a standalone information processing unit capable of processing received stimuli in parallel much like its biological counterparts. These neurons are organized into pre-defined layers of ANN. The first layer is the input layer while last layer of ANN is the output layer. In between these two layers are a number of connected hidden layers that act like a series of transformation functions, which slowly maps input to the output of system. We have designed an ANN with four layers one input, two hidden layers and an output layer. Each hidden layer contains five nodes and log sigmoid transfer function while output layer is designed to have two nodes with soft max transfer function as shown in figure3.

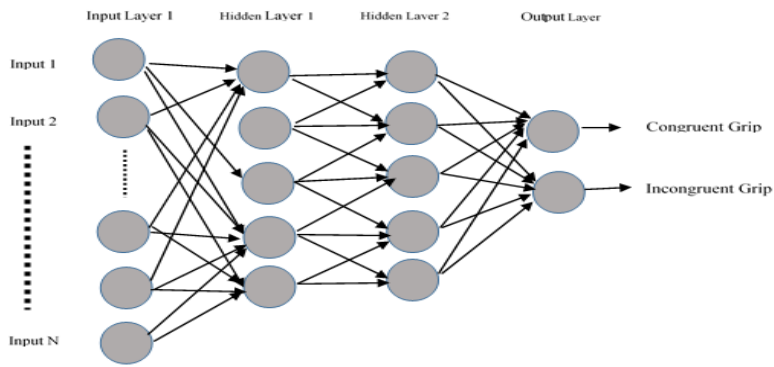

Fig. 3. General Architecture of Artificial Neural Network.

The next section focuses on interpretation of results obtained during and after the experiment.

## 3 RESULTS AND DISCUSSION

The ERD/ERS calculations for both events clearly showed that there is a significant deflection in the power of alpha band wave with respect to the type of the handgrip. This deflection is due to the differences in the power spectrum corresponding to EEG signal associated with each event. For our purpose, power spectral density was calculated for each event across epochs and used as a feature of interest for classification purpose. These features are then fed to an ANN for classifying the event as congruent handgrip or incongruent handgrip.

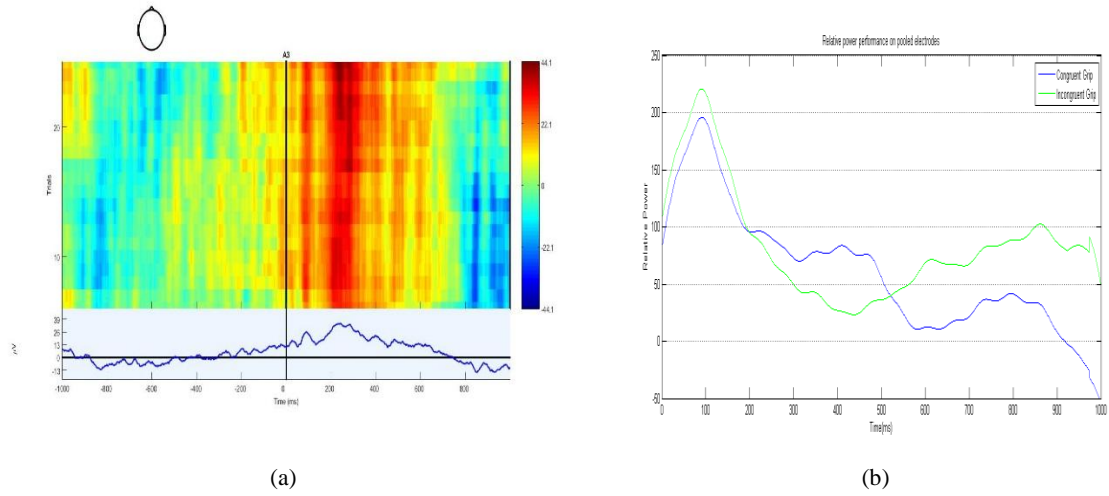

(a)            (b)

Fig. 4. (a) Averaged Power Spectrum for Alpha band showing high power at the time of event occurrence

(b) Relative Power Performance on pooled electrodes for both events

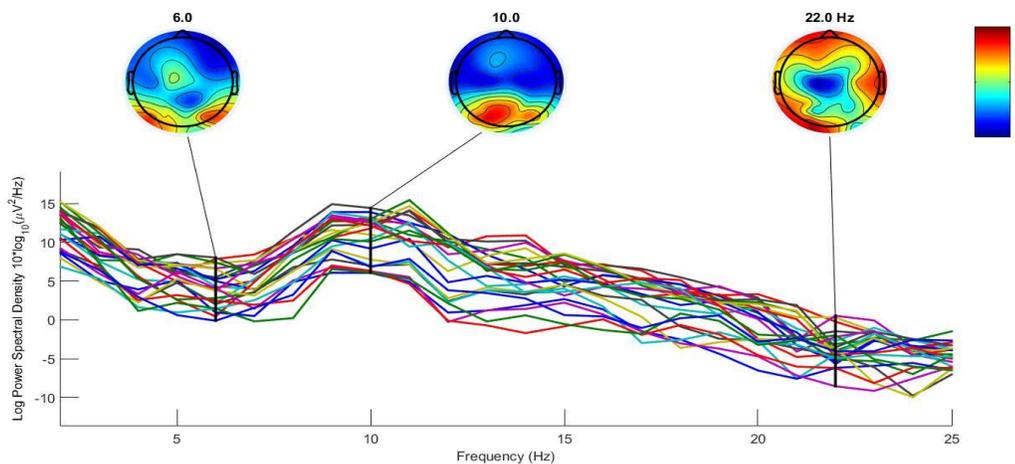

Fig. 5. (a) Power spectral density across all channels showing a prominent peak for frequency range 8-12 Hz (alpha band)

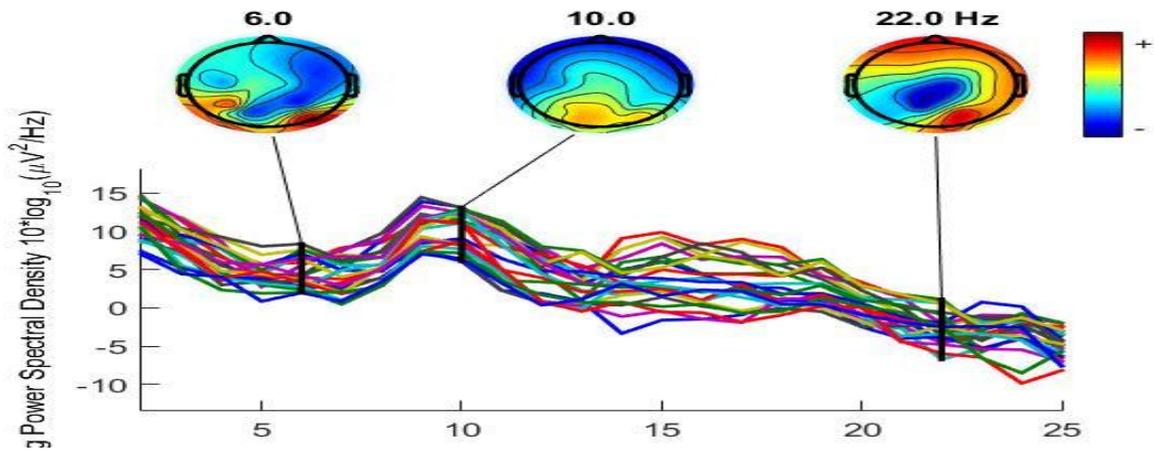

Fig. 5. (b) Power spectral density across all channels showing a prominent peak for frequency range 8-12 Hz (alpha band)

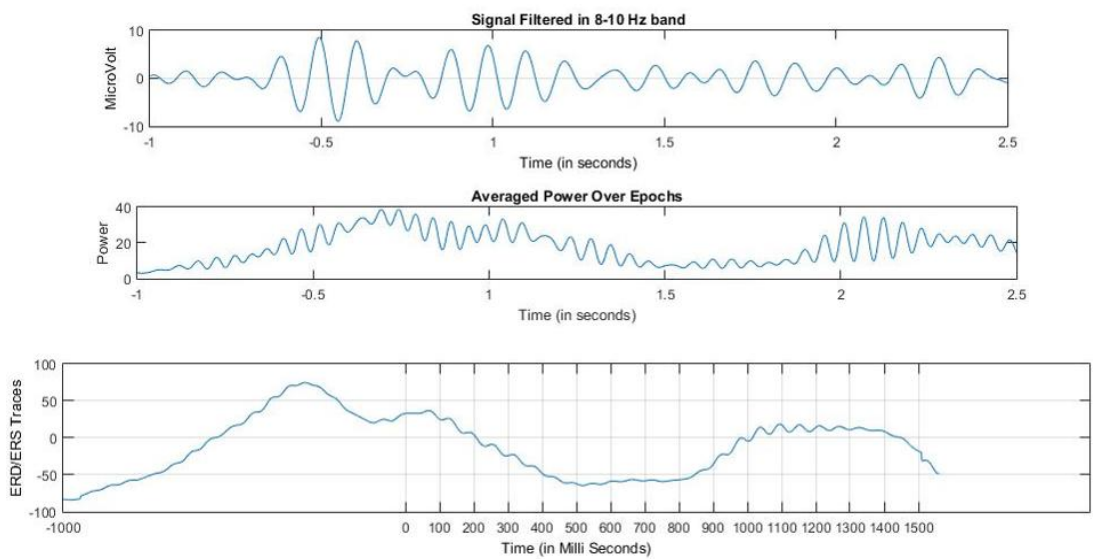

Fig. 6.

- ERD/ERS calculation for each sample point
- ERD/ERS= ((Power in Activity Period for each point- mean Power in baseline period) / - mean Power in baseline period*100;

XX:9

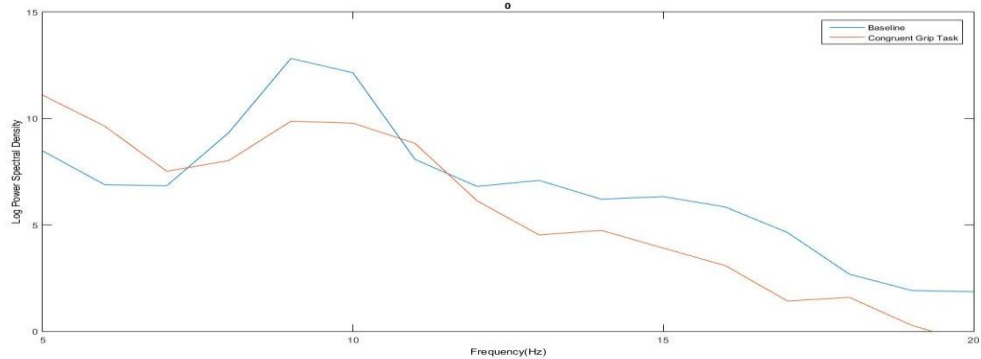

Fig. 7. Log Power spectral density showing difference in peak power of baseline event (Rest period) and congruent grip task

## 4  CONCLUSIONS

In this work a computer aided system was proposed to automatically classify subjects having correct and incorrect grip over objects of interest based on EEG signal analysis. The method involves extraction of features from different EEG sub bands obtained by applying DWT followed by an ANN classifier to classify incoming signal data into either congruent or incongruent grip responses. The proposed system can be used to develop brain machine interfaces for neurorehabilitation purpose.